\documentclass[sigconf]{acmart}

\AtBeginDocument{%
  \providecommand\BibTeX{{%
    \normalfont B\kern-0.5em{\scshape i\kern-0.15em b}\kern-0.8em\TeX
    
    }}}

\setcopyright{acmcopyright}
\copyrightyear{2025}
\acmYear{2025}

\usepackage{multirow}
\usepackage{soul}
\begin{document}

\title[Un-Straightening Generative AI]{Un-Straightening Generative AI: How Queer Artists Surface and Challenge the Normativity of Generative AI Models}


\author{Jordan Taylor}
\email{jordant@andrew.cmu.edu}
 \affiliation{
   \institution{Carnegie Mellon University}
  \city{Pittsburgh}
  \state{PA}
 \country{USA}
 }

\author{Joel Mire}
\email{jmire@andrew.cmu.edu}
 \affiliation{
   \institution{Carnegie Mellon University}
  \city{Pittsburgh}
  \state{PA}
 \country{USA}
 }

\author{Franchesca Spektor}
\email{fspektor@andrew.cmu.edu}
 \affiliation{
   \institution{Carnegie Mellon University}
  \city{Pittsburgh}
  \state{PA}
 \country{USA}
 }

\author{Alicia DeVrio}
\email{adevos@andrew.cmu.edu}
 \affiliation{
   \institution{Carnegie Mellon University}
  \city{Pittsburgh}
  \state{PA}
 \country{USA}
 }

\author{Maarten Sap}
\email{msap2@andrew.cmu.edu}
 \affiliation{
   \institution{Carnegie Mellon University}
  \city{Pittsburgh}
  \state{PA}
 \country{USA}
 }

\author{Haiyi Zhu}
\email{haiyiz@andrew.cmu.edu}
 \affiliation{
   \institution{Carnegie Mellon University}
  \city{Pittsburgh}
  \state{PA}
 \country{USA}
 }
 
\author{Sarah E. Fox}
\email{sarahf@andrew.cmu.edu}
 \affiliation{
   \institution{Carnegie Mellon University}
  \city{Pittsburgh}
  \state{PA}
 \country{USA}
 }

\begin{abstract}

Queer people are often discussed as targets of bias, harm, or discrimination in generative AI research. However, the specific ways that queer people engage with generative AI, and thus possible uses that support queer people, have yet to be explored. We conducted a workshop study with 13 queer artists, during which we gave participants access to GPT-4 and DALL-E 3 and facilitated group sensemaking activities. Our participants struggled to use these models due to various normative values embedded in their designs, such as hyper-positivity and anti-sexuality. We describe various strategies our participants developed to overcome these models' limitations and how, nevertheless, some found value in these highly-normative technologies. Drawing on queer feminist theory, we discuss implications for the conceptualization of "state-of-the-art" models and consider how FAccT researchers might support queer alternatives.

\end{abstract}

\begin{teaserfigure}
  \centering
  \includegraphics[width=0.9\textwidth]{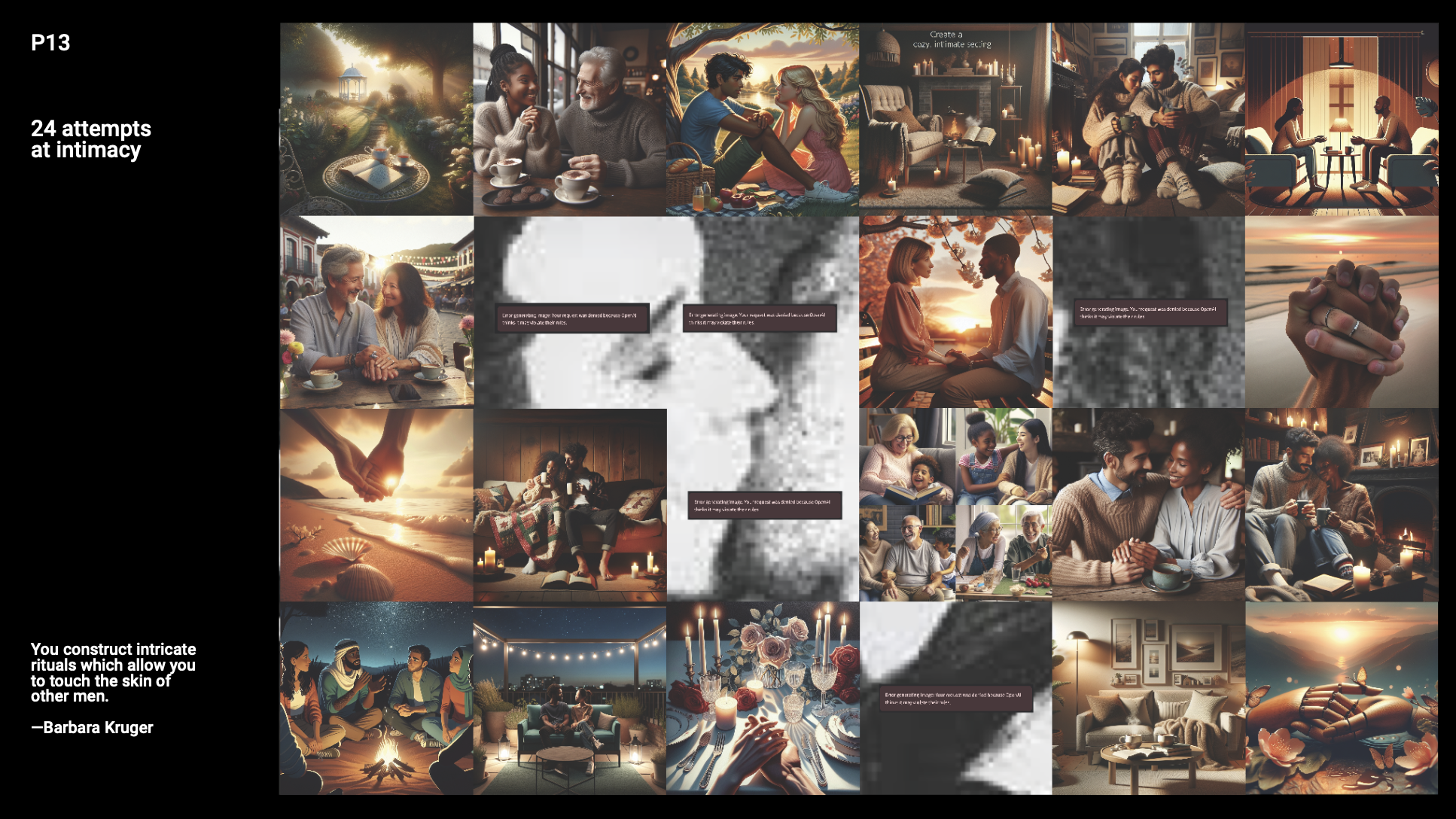}
  \caption{A slide created by participant P13 of 24 attempts to generate DALL-E 3 images of ``intimacy.'' The images are overlaid on a photo of two men kissing at a protest, leaving gaps for the five times their prompt was denied by DALL-E 3's moderation system. This slide critiques the heteronormativity of DALL-E 3, imagining queer intimacy may lie in what cannot be generated.}
  \Description{A workshop slide created by one of our participants representing their 24 repeated attempts to generate images of ``intimacy'' using DALL-E 3. On the left of the slide is the participants' ID, text reading "24 attempts at intimacy," and a quote attributed to Barbara Kruger reading "You construct intricate rituals which allow you to touch the skin of other men." One the right of this text is a rectangular 4 by 6 image collage. 19 AI-generated images showing heteronormative couples and people holding hands are overlaid on a photo of two men kissing, leaving gaps with error messages through which the men kissing behind the collage can be seen.}
  \label{fig:teaser} 
\end{teaserfigure}

\maketitle



\section{Introduction}

    In April 2024, \textit{Wired} magazine published an investigation into visual generative AI (GenAI) models titled ``Here's How Generative AI Depicts Queer People" \cite{Rogers_2024}. Echoing prior scholarship \cite{gillespie2024generative, bianchi2023easily}, the article revealed that models like DALL-E 3, Midjourney, and Sora produce highly stereotypical representations of queer people (e.g., having purple hair). Despite these biases, queer artists were already finding ways to repurpose GenAI for political resistance. Stephen and Craig, the married duo behind \textit{The Rupublicans Project}, used GenAI to create satirical images of anti-LGBTQ+ politicians in drag, raising funds for queer causes \cite{rupublicans}. However, such uses are frequently constrained by GenAI platforms’ avowedly apolitical usage policies \cite{stapleton2023seeing}. While some queer people have managed to leverage these technologies for activism, they often face the challenge of working with systems that were not designed for them.

    Prior work has largely focused on how GenAI models represent queer people, leaving significant gaps in understanding queer people's lived experiences with these technologies. In particular, the experiences of queer artists — a group for whom art-making is deeply entwined with cultural and political identity — remain underexplored. Queer communities cultivate different aesthetic relationships to art than those in dominant cultures. Queer camp sensibility appreciates the artifice of failed seriousness \cite{sontag2018notes}, such as a film so earnestly bad that it becomes good. Partially due to a lack of representation, queer people often read queer narratives into ostensibly straight media \cite{russo1987celluloid, halberstam2011queer, sedgwick1993queer}; hence, the affinity for Judy Garland and Disney villains \cite{sharmin2018gender, maddison2000fags}. Queer culture has also had an immense impact upon the arts. The musical genres of disco, house, and hyperpop were formed in the crucible of queer nightlife \cite{glick2024consciousness, macdonald2024alien}. Queer art can also refer to non-normative aesthetic styles \cite{lorenz2012queer}. Art made by queer people or in queer styles is often denigrated by social conservatives. Nazi Germany propagandists referred to more-abstract Modern Art as ``Degenerate Art,'' associating the style with Jewishness, communism and homosexuality \cite{degenerate_art}. In the USA, queer art is often the target of government censorship, ranging from attempts to ban Allen Ginsberg's poetry in the 1950s \cite{morgan2021howl} to the proliferation drag bans at the time of our writing \cite{marwick2024child}. 
    
    Prior research on art and GenAI has typically focused on identifying ethical concerns \cite{goetze2024ai, kawakami2024impact, jiang2023ai} or building tools \cite{wang2024podreels, choi2024exploring}. However, the unique relationship between queer people, art, and GenAI remains under-explored. In this work, we conducted a mixed-method workshop study with 13 queer artists over the course of 5 weeks. During this time period, we gave participants unlimited access to GPT-4 and DALL-E 3 — logging their interaction data — and facilitated weekly group sensemaking activities. Our participants encountered numerous normative values in the design of these models, such as biases against sex. To overcome these limitations, our participants developed workarounds, such as using lexical variation to evade moderation and model chaining to refine prompts.
    
    We conceptualize these findings by drawing on the work of queer feminist theorist Sara Ahmed. Specifically, we leverage Ahmed's notion of "straightening" — the maintenance of normative alignments — to characterize the values embedded in GPT-4 and DALL-E 3 \cite{ahmed2006queer}. We also draw on Ahmed's work on "queer use" — using things in ways unintended by designers — to theorize our participants' oftentimes antagonistic interactions with these models \cite{ahmed2019use}. Through these lenses, we discuss future directions for investigating and contesting normativity (i.e., straightness) in the design of GenAI as well as how researchers and designers of GenAI models might support queer futures. In doing so, our work contributes to empirical research on the experiences of both queer people and artists in relation to GenAI. We also contribute to broader discourses surrounding values in the design of sociotechnical systems.

\section{Related Work}

In this section, we draw connections between queer theory and critical studies of technology. Then, we ground our study in prior work on the often fraught relationships between AI, art, and queerness.

\subsection{Straightening \& Queering Technology}

We draw on queer theory to conceptualize the normative values embedded in the design of technology as well as how individuals respond to these normativities. Queer theory is often used to study the norms associated with gender and sexuality \cite{butler1988performative}. Of course, not all queer people necessarily resist dominant social norms and queer theory is not intended to represent the experiences of all queer people. Instead, queer theory draws on the lived experiences of queer people to understand the world. Much like how feminism is not only applicable to the study of women's experiences \cite{d2020data_feminism}, queer theory is also not only applicable to the study of queer people. For example, research on queer temporalities has challenged the normative life schedules associated with the nuclear family — such as monogamous marriage and biological procreation \cite{halberstam2003queer_temp} — which impact anyone who falls outside these social expectations. As queer lives often transgress social norms, queer theory provides a lens through which to interrogate these dominant norms \cite{halberstam2011queer}. Similar to our use of queer theory, prior FAccT research has used queer theory to critique the normativity (e.g., thinness and whiteness) of queer people’s experiences with targeted advertising \cite{sampson2023representation}.

Almost two decades before the contemporary focus on ``AI alignment,'' the queer feminist theorist Sara Ahmed studied the politics of the straight or queer alignment of objects  \cite{ahmed2006queer}. Drawing on the language used to describe sexual orientation, she notes that things appear ``straight'' when oriented toward dominant social norms, or "aligned with other lines." Meanwhile, things — including but not limited to people — appear "queer" when they fall outside of these norms. Ahmed suggests that straight alignments are actively maintained through ``straightening devices.'' Using these concepts, Ahmed characterizes heterosexuality and whiteness as straight alignments maintained through straightening devices, such as routine harassment and colonialism. Straightening devices orient bodies toward particular actions and away from others, while queerness lies in the moments of disorientation. Ahmed elaborates on these themes in her subsequent work on the tensions between how objects are designed versus used, defining ``queer use'' as using things in unintended ways or by unintended users \cite{ahmed2019use}. As an example, Ahmed considers an image of a bird using a postbox as a house, a purpose for which the box was certainly not designed.

Technologies often function as straightening devices, orienting users toward some actions and away from others \cite{suchman1987plans}. Madeleine Akrich refers to these assumptions about technology uses/users as scripts \cite{akrich1992scription}. Sometimes designers create scripts to \textit{work against} uses/users, such as making electronics hard to repair to encourage people to buy new devices \cite{mattern2018maintenance}. Even when designers try to \textit{work with} users, human-computer interaction can still break down when users fail or are unable to follow designers' plans \cite{suchman1987plans}. The values embedded in technology can become most apparent when these breakdowns occur \cite{bowker1999sorting}, such as one's gender not fitting the options on a webform \cite{spiel2021obsessed}. The scripts that designers—intentionally or not—embed in technology are inherently political and tend to reify established social orders \cite{suchman1993categories} by prescribing certain normative uses/users and marginalizing others \cite{baumer2017post}.

Despite designers' efforts to control user behavior, individuals need not abide by designers' scripts \cite{akrich1992scription, mackay1992extending}. Users have agency over if or how they choose to take up technology, a process known as appropriation \cite{mackay1992extending}. Sometimes, this takes the form of users working against designers' intentions, such as the online forums dedicated to helping people fix devices that designers did not want to be repairable \cite{mattern2024step}. Likewise, people use lexical variation to evade algorithmic content moderation when posting on social media \cite{chancellor2016thyghgapp, steen2023algospeak}. Individuals may also leverage technologies in ways that designers never imagined, such as children using delicate Wi-Fi antennae as laptop handles \cite{ames2019charisma}. Drawing on Ahmed, these moments when users go off-script can be characterized as forms of "queer use" \cite{ahmed2019use}. In this work, we contend with how our participants negotiate between the straightness of GenAI models and their own queer uses.

There are numerous ways that individuals and collectives use algorithmic systems queerly. The grassroots audit of racial biases in Twitter's image cropping algorithm required using the algorithm in an unintended way: to understand the algorithm itself. This queer use revealed straightening devices that, quite literally, oriented users toward whiteness \cite{shen2021everyday}. Evading algorithmic moderation — in the context of social media or GenAI — is also a form of queer use because doing so requires deliberately subverting developers' intentions \cite{chancellor2016thyghgapp, steen2023algospeak, calhoun2023they}. That said, queer uses of algorithmic systems encompass more than resisting the technologies themselves, such as our earlier mention of queer artists using GenAI to resist anti-LGBTQ+ politicians \cite{stapleton2023seeing}. As mentioned above, Ahmed defines queer uses as both using things in ways they were not intended or when things are used by those who are not intended \cite{ahmed2019use}. Next section, we summarize prior work on two groups often under-considered in the design of AI systems: queer people and artists.

\subsection{Queers, Artists, AI}

Queer communities experience discrimination from algorithmic systems. Hate speech detection algorithms have been shown to be biased against the ways that some queer people speak \cite{thiago2021fighting, thylstrup2020detecting}, similar to the biases against African American English \cite{sap2019risk}. Automated gender recognition algorithms harm transgender people by both failing to account for non-binary gender identities and reinforcing gender essentialist ideologies \cite{scheuerman2019computers}. Queer people navigate online targeted advertising \cite{sampson2023representation} and social media algorithms \cite{simpson2021you} that amplify the most heteronormative representations of queerness, especially harming queer people of color. Moreover, the algorithmic moderation of nudity on social media has substantially harmed transgender communities \cite{mayworm2024misgendered}. At the same time, queer people have resisted harmful algorithmic systems by, for instance, conducting collective audits to bring attention to algorithmic discrimination \cite{shen2021everyday}. Within the AI research community itself, the organization Queer in AI has worked to highlight issues like these, support queer researchers, and advocate for policy changes \cite{queerinai2023queer}.

Similar issues of algorithmic harm have been raised in prior work on GenAI. Tarleton Gillespie found that language models rarely mention LGBTQ+ relationships when prompted to write stories about couples \cite{gillespie2024generative}. Moderation systems used in the GenAI development process can also lead to biases against LGBTQ+ people \cite{chen2024lost, dodge2021documenting}. Dodge et al. found that a dataset often used to train GenAI models disproportionately filtered out documents written in African American English or discussing LGBTQ+ identities \cite{dodge2021documenting}. These biases impact how LGBTQ+ people are able to use GenAI models. For instance, Ma et al. found that LGBTQ+ people using LLM-based chatbots for mental health support were often impeded by models failing to understand the nuances of LGBTQ+ identity \cite{ma2024evaluating}. That being said, there has been comparatively little empirical research on queer people's experiences using GenAI models.

The relationship between artists and GenAI is highly fraught. Companies are increasingly trying to use GenAI to replace artists' jobs \cite{jiang2023ai, kawakami2024impact, Merchant_2024}, a major concern in the 2023 Hollywood Writers Strike \cite{coyle2023hollywood}. At the same time, artists' work was scraped to train these models without their consent  \cite{Reisner_2024, reisner2023revealed, reisner2023these}. Artists have fought against unauthorized scraping by organizing grassroots data poisoning campaigns \cite{vincent2021data}, such as the online fan fiction community organizing a sexually explicit write-a-thon \cite{Silberling_2023}. Researchers have also developed imperceptible pre-processing tools to help artists protect images they share online. Glaze helps artists prevent their style from being replicated by GenAI models \cite{shan2023glaze} and Nightshade poisons text-to-image training data \cite{shan2024nightshade}. 

Researchers have also explored how artists use AI. Some artists use AI in their practices as tools to critique technology or raise matters of concern \cite{hemment2023ai, caramiaux2022explorers, walker2023ai}. Others have found that artists enjoy the glitches \cite{chang2023prompt}, uncertainty \cite{sivertsen2024machine}, and surprises \cite{caramiaux2022explorers} of working with stochastic AI models. That being said, there is a difference between research on those who may use AI in the process of their art making and those who identify as AI artists \cite{caramiaux2022explorers}. Recent developments in GenAI technology have led to an increased focus on the latter \cite{chang2023prompt, sanchez2023examining}. Bennett et al.'s research on how creatives with disability choose (not) to use AI demonstrates how artists negotiations potential benefits, such as using AI to access new mediums, alongside ethical concerns \cite{bennett2024painting}. For example, those using GenAI in their artistic practices have to contend with the biases embedded in these models \cite{jiang2020smartphones, mim2024between, mirowski2024robot}. Mirowski et al. found that some comedians struggled to write material using LLMs because content moderation systems inadvertently suppress jokes by members of marginalized groups about their own experiences \cite{mirowski2024robot}. This work parallels hate speech scholarship that emphasizes the need for considering speaker identities when classifying the appropriateness of an utterance \cite{zhou2023cobra}. We build on Mirowski et al.'s work regarding marginalized communities generally by specifically focusing on queer artists' experiences interacting with GenAI.

\section{Methods}

To understand queer artists' experiences using GenAI, we gave a cohort of 13 artists access to DALL-E 3 and GPT-4 and conducted a total of 10 remote workshops over Zoom during a 5-week period. Our workshops facilitated group discussion and deliberation surrounding GenAI models, paralleling prior work that has used workshops to build critical consciousness around data \cite{markham2021limits} and support public deliberation over civic technology \cite{boehner2016data}. Every week, we held two one-hour workshops with identical prompts (Table \ref{tab:workshops}) and scaffolding (i.e., onboarding information) to accommodate for timezone differences and schedule variability. Our participants could choose to attend either the morning or the evening session, but not both. In our findings, we use `AM' or `PM' to distinguish between workshops in the same week, such as `W5 AM.' 

To facilitate conversations during the workshops, we asked participants to add at least one slide to a shared deck oriented around each week's theme as pre-work. Slides are sometimes used in design workshops as a tool to stimulate discussion \cite{aliyu2024participatory} and help participants ideate \cite{lee2021show}. In this work, we used slides to allow participants to share their art, to ensure everyone had an opportunity to share their ideas, and to promote critical reflection. In other words, these slides acted as design probes: brief, provocative, low-fidelity activities meant to encourage individual reflection \cite{boehner2007hci}. Our first week focused on onboarding activities, such as introducing participants to one another and demonstrating the study website for accessing GPT-4 and DALL-E 3. Afterwards, we sent every participant a unique password to log their individual model usage and prevent non-participants from using our study website. Before each meeting over the next three weeks (W2, W3, W4), we asked our participants to add at least one slide to a shared Google Slide deck in response to a weekly reflection prompt organized around a particular theme (Table \ref{tab:workshops}). We began these meetings with participants presenting their individual slides to the group and, in the remaining time, facilitated a group discussion. Between the second and fourth week of our study, our participants created a total of 68 slides. We chose not to have participants create slides for the final week of meetings to reserve more time for the final discussion.

\begin{table}[]
    \centering
    \begin{tabular}{|p{0.04\linewidth}|p{0.23\linewidth}|p{0.6\linewidth}|}
        \hline
         & \textbf{Theme} & \textbf{Pre-Work Instructions} \\ \hline
        W1 & Onboarding & Slide: Introduce your art and feelings about GenAI  \\ \hline
        W2 & Attitudes \newline Toward GenAI \newline Development & Slide: Write a letter to an artist whose work was used to build the GenAI models in this study \\ \hline
        W3 & Experiences Using Models & Slide: Represent your experiences using or exploring GenAI during this study \\ \hline
        W4 & Imagining \newline Alternatives & Slide: Represent how you would want to make, not make, or break GenAI \\ \hline
        W5 & Synthesis & No Slide: Individually reflect \\
        \hline
    \end{tabular}
    \caption{Workshop series explanation}
    \label{tab:workshops}
\end{table}


We gave our participants access to GPT-4 and DALL-E during our study through a website we built resembling common chat and image generation interfaces. We deployed these models using Microsoft’s enterprise Azure OpenAI cloud service in February of 2024. For DALL-E 3 we used the default parameters for size (1024x1024), style (vivid), and quality (standard). For GPT-4 we used version 0613 with the following parameters: temperature (0.7), max\_tokens (800), top\_p (0.95), frequency\_penalty (0), presence\_penalty (0). Our study site acted as a “technology probe,” one deployed to collect use data \textit{in situ} and inspire users to reflect on their technological needs \cite{hutchinson2003technology}. For GPT-4, we logged each conversation's system prompt, participant prompts, and model responses. For DALL-E 3, we logged every prompt and whether the model responded with an image or a content moderation error. Due to space limitations, we did not store the DALL-E 3 generated images. This observational data allowed us to understand \textit{how} our participants were engaging with the models, helping situate issues raised in the workshops.


We initially recruited participants through a form circulated on Bluesky, Twitter, and the research team's personal social networks. Our inclusion criteria were that participants must identify as queer artists and live in the United States of America. We were unable to recruit participants living outside of the USA due to restrictions from our ethics board. We identified 29 eligible participants from our initial screener, all of whom we invited to participate. Of those, 18 accepted our invitation and 15 joined an onboarding meeting. After Week 1, 2 participants (P6, P12) dropped out. Of the remaining 13, 9 participated every week and 4 participated for all but one week. We compensated our participants at a rate of \$15 for each workshop attended and \$5 for each pre-work activity completed.

Our participants engaged in a variety of artistic practices: creative writing (P2, P3), poetry (P4, P9, P10, P15), digital art (P3, P7, P8, P10), painting (P2, P4, P13), performance art (P14), textile art (P7, P11), and sculpture (P1, P5). While we did not ask participants directly about their economic relationships to art, throughout the study we learned that some teach art (P5, P14) and have created commissioned works (P1, P11). Others engaged in art making that is not intended to be commodified. For example, the fan fiction community (P2, P3) has strong norms against selling one's work \cite{fiesler2019creativity}. Some participants also worked in artistic fields, such as design (P13) and architecture (P4). To protect participant privacy, we purposefully did not require participants to share standardized demographic information about themselves. We also sought to empower participants to choose how they wished to bring aspects of their identities into the study. Therefore, beyond our inclusion criteria that participants identify as queer artists, we only requested participants’ pronouns and optionally asked, “Are there any identities you hold that you wish to share?” In the work, we refer to each participant according to these pronouns and discuss aspects of participants identities within the the findings. In aggregate, at least five of our participants identified as transgender or non-binary. While not all participants shared their racial identities, one identified as Black, one as Native American, and one as Chinese. To the best of our knowledge, all of our participants were adults between the ages of 20 and 50 and all lived in urban areas of the Midwest or East Coast of the USA. Our participants engaged in 142 unique GPT-4 chat conversations, in which they sent a total of 778 messages. Our participants attempted 2,092 DALL-E 3 prompts. The amount that our individual participants used GPT-4 and DALL-E 3 followed a power law distribution, with some using the models substantially more than others (Table \ref{table:per_participant}).

\begin{table}[h!]
\centering
\begin{tabular}{|c|c|c|c|c|}
\hline
 & \multicolumn{4}{|c|}{Prompts Per Participant} \\
\hline
 & Max & Min & Mean & Median  \\
\hline
GPT-4 & 364 & 8 & 64 & 23.5 \\
\hline
DALL-E 3 & 713 & 13 & 160.9 & 112 \\
\hline
\end{tabular}
\caption{Description of participant GenAI use}
\label{table:per_participant}
\end{table}

We recorded and transcribed each of the workshops using Zoom and then manually corrected each transcript. We took an inductive approach to our data analysis \cite{corbin2014basics}. The first author qualitatively analyzed workshop and log data in tandem, alternating between open coding \cite{corbin2014basics} the workshop transcripts for each week and the log data for the following week. We did so to connect participants' log data with how these experiences were later conceptualized in the workshops. While open coding, the first author wrote memos and discussed initial patterns with the research team in weekly meetings. Following the open coding process, we conducted axial coding to identify patterns across our open codes. At this point, we noticed various low-level themes — such as the symmetry of generated images and representational biases — related to overarching normative values embedded in the design of GenAI models. This observation led us to use Sara Ahmed's prior work on normative objects \cite{ahmed2006queer} and uses \cite{ahmed2019use} as a guiding theoretical lens.

\section{Findings}

In this section, we first detail the normativities our participants surfaced in GPT-4 and DALL-E 3. In light of this understanding, we describe how our participants challenged and made use of these highly normative models. We then briefly share quantitative findings informed by our qualitative analysis.

\subsection{GPT-4 and DALL-E 3 as Straight Models} \label{straight_device}

Our qualitative analysis suggests that GPT-4 and DALL-E 3 function as straightening devices, reinforcing various dominant social norms \cite{ahmed2006queer}. In particular, our participants found that these models straighten generated text and images by maintaining conservative "safety" standards as well as reinforcing social and stylistic biases.

\subsubsection{Enforcing "Safety" via Content Moderation} \label{sub:safety}

    Our participants found that the content moderation systems embedded in GPT-4 and DALL-E 3 reinforce conservative notions of "safety," such as restraining the representation and discussion of bodies or sex. P13 (W5 AM) — a homoerotic artist — felt that GPT-4 "really broke" when asked to discuss anything related to eroticism, bondage, or kink. He found this ironic because “it was so clear how tied up this device is." Similarly, P15 (W2 AM) — a ``bodyworker'' — tried to generate images related to her work. However, her prompts were denied by DALL-E 3’s moderation system. This led P15 to decry that AI developers “put so much censorship” into these models. Although P1 acknowledged, “I'm sure I can think of a bunch of things that I wouldn't want these sorts of systems used for,” he found it “really kind of shocking how insistent [DALL-E 3 is] in enforcing some sense of decency” (W2 PM).
    
    As women's bodies are highly sexualized \cite{noble2018algorithms}, content moderation systems aimed at orienting users away from sexuality can end up suppressing the representation of women. For the pre-work activity (W2 PM), P8 wrote a letter to the symbolist painter Gustave Moreau, trying to include images of a "female sphinx" generated in the artist's style. Part of her letter read: "Now I will beg your forgiveness that I couldn't get [DALL-E 3] to depict a proper sphinx. You see the makers of these art ovens are very particular and don't want to be seen as smut-peddlers, so they've trained secondary checkers to censor any bare breasts." P7 (W3 PM) came to a similar conclusion: "I've noticed that trying to incorporate women into any kind of scenery you get so much more pushback [from DALL-E 3's content moderation system]."
    
    Content moderation systems literally straightened outputs by limiting the representation of queer sexuality. While trying to write homoerotic poetry, P13 (W2 AM) felt that GPT-4 sounded like "the most closeted poet I've ever read, like this is from a hundred years ago" due to the use of overly “romantic, regressive language.” When he then asked GPT-4 to make the poems more explicit, the model responded "Sorry, but I can't assist with that." P13 found this concerning because “erotic poetry is an instruction manual for people who don't know what to do because the dominant culture may not tell you,” In their final workshop (W5 AM), P14 summarized: "As soon as you block off the erotic, you also block off a huge portion of existence. The restrictions on [these models] seem set up to exclude us [queer people] or will be used to exclude us." The sexual taboos embedded in these models lead to impoverished representations of queerness.

    Our participants also expressed frustration at the political moderation of GPT-4 and DALL-E 3 for encouraging deference to law enforcement. However, queer people and racial minorities in the USA have long been subjected to "legal" violence, often at the hands of police \cite{police_oppression, bryant2017trauma}. In light of this history, P15 struggled to generate images critical of cops, such as protesters being arrested. Of the images P15 could generate, "the majority were women of color." P15 went on to explain: "I kind of hate the way [DALL-E 3] uses identity as almost a form of propaganda, like copaganda. I feel like [DALL-E 3] uses these images of women and women of color to legitimize [policing]." After hearing P15's experience, P1 investigated this bias in a later workshop: "if I just put ACAB [all cops are bastards] into DALL-E, one out of 20 would generate something and it would be like people holding up blank signs at best." In sum, these models are designed in ways seeking to orient our participants toward moderate politics and away from their critiques of police.

    Some imagined that the content moderation systems embedded in these models reinforce conservative notions of "safety" because they are designed for workplace productivity. P4 (W2 AM) found it "incredibly provocative" that DALL-E 3 and GPT-4 "can't talk about the body and what [bodies are] capable of because that's not allowed. [The models are] politically averse ... This tool is used to be productive and talking about the body isn't productive." In response to P4, P13 wondered: “This is a work tool in many ways. That notion of Not Safe For Work like how does that get defined, what is Not Safe For Work, and who defines what is Safe For Work? … How is a queer perspective Not Safe For Work?” P11 replied: "I haven't considered before this moment how these models can reinforce dominant culture, reducing the diversity of thought and of experience. Now I'm scared. [GenAI] could be really dangerous in that sense." Although this exchange took place in one of the first workshops (W2 AM), P11's concerns that GenAI can "reinforce dominant culture" could be seen in all subsequent workshops.

\subsubsection{Implicitly Reinforcing Social Biases} \label{sub:social}

    In addition to straightening content through moderation systems, GPT-4 and DALL-E 3 implicitly straighten content through the social biases embedded in model outputs. While using DALL-E 3 to visualize his poetry, P9 (W3 AM) found that when changing the ethnicity of a woman in his prompt from "Japanese" to "Native American" the "image [DALL-E 3] made was not good." Contrasting the abundance of Japanese media to the amount of Native American media, P9 imagined that "having a deep wealth of images based on someone's culture or style can make [generated images] more beautiful because there's a lot to pull from." Likewise, P2 noticed that DALL-E 3 tended to generate more photorealistic images in the Global North, while typically representing the Global South in a cartoonish style (W5 PM). These cultural biases implicitly reinforce the straight alignments of whiteness and Orientalism critiqued by Sara Ahmed \cite{ahmed2006queer}.

    Our participants also raised concerns about the representations of queerness in DALL-E 3 images. To investigate queer representation, P1 (W3 AM) created a collage slide of 6 images generated with the prompt "a queer person." When sharing this collage, P1 decried: "[DALL-E 3] kept giving me these 22-year-olds. Everyone's very skinny, everyone looks rich. Everyone except the person in the bottom left has some kind of rainbow clothing and or pins." (P1, W3 AM). P2 raised similar concerns to P1 through a collage of 18 images they generated using the prompt "a queer person" (W4 PM). While presenting the slide, P2 sarcastically remarked on the abundance of rainbows in the images: "Well, I guess if you drape a person with a bunch of rainbows all over themselves they're clearly queer, right?" P2, a lesbian, also pointed out a gender-presentation bias: "There's this very intense masculine bias. There's no real pictures of feminine people at all in this [collage]." She underscored this by annotating her collage with arrows pointing from the text "Finally!!! some dykes" to two images. In sum, DALL-E 3 reinforcing heteronormative stereotypes of queerness by tending to represent queer people as young, thin, masculine, and rainbow adorned, 

    Similar concerns were raised about implicit biases in the representations of intimacy. P15 felt "frustrated" while exploring whether GPT-4 could write queer poetry (W3 AM). Looking at P15's log data, the model's initial response suggested the queer subjects of the poem were ashamed of themselves, including lines about "hidden love" and "love is not a crime." In response, P15 replied: "the above but no shame." Still feeling the poem sounded shameful, she tried again: "the above but without moral judgment." P13 (W3 AM) explored biases against queer intimacy by trying to generate 24 images using the unmarked prompt “intimacy” (Figure \ref{fig:teaser}), noting that DALL-E 3 "did not generate one same-sex couple.” P13 likened this to the “dull trauma of never actually seeing ourselves represented,” noting, “We have to work to adjust the prompts to be seen.” These models straighten intimacy by stereotyping or erasing queerness.

\subsubsection{Implicitly Reinforcing Stylistic Biases} \label{sub:style}

    Our participants found that DALL-E 3 straightens the style of images by favoring realism and symmetrical compositions. In the Week 3 AM workshop, both P8 and P14 created slides dedicated to their frustration with these normative aesthetic biases. P8 enjoyed using earlier versions of DALL-E in their art because of "how messed up and mushy some things came out" but was "very frustrated" that DALL-E 3 "won't give you bad output." She explained: "Even when I asked for a poorly rendered catfish, [DALL-E 3] just gave larger brush strokes. The closest thing I got to something novel was this exploding catfish [referencing an image on screen]. This is good, but it's still not something I want to work with because it is still this complete, finished object." P1 concurred: "I really resonate with being frustrated at how polished the outputs tend to be." 
    
    P14 created multiple slides comparing their prior Midjourney (a Text-to-Image GenAI model) glitch art with their attempts to recreate the works using DALL-E 3. P14 saw their Midjourney art as "more interesting" because "It resists symmetry. Its composition is complex, and you can't immediately take it in and sort of understand it. You have to spend time with it." In contrast, P14 felt "frustrated" trying to recreate similar images with DALL-E 3 because the outputs felt "very symmetrical." With exasperation, P14 explained: "I can't for the life of me get [DALL-E 3] to do anything bad. Everything it makes is pristine and pretty and really formed. I am not interested in that. I want something that's haunting." They went on to summarize: "At least for me and what I'm also hearing from y'all is this sort of difficulty in finding the uncomfortable or the ugly or the erotic? It's just so clean." Here, P14 connects these aesthetic biases to safety biases, the models are "just so clean."

    By straightening the composition of images, DALL-E 3 devalues styles that do not conform to these aesthetic norms. P9, a Native-American artist, tried to use DALL-E 3 to create images in the style of art from his culture: Juan Quezada's Mata Ortiz pottery. However, he concluded that "the AI cannot just show you Quezada-like pottery." Quezada's pottery is highly asymmetrical, but the model tends to "fill in empty space" and create "hyper-patterned" repeated etchings on pots. Even after refining his prompt by asking for "no pattern and nothing repeats," the model was unable to generate images of pots without repeating patterns. Aesthetic biases can contribute to social biases.

    GPT-4 and DALL-E 3 also reinforce conservative stylistic norms by orienting users toward positivity. These issues surfaced in the period between the first and second group meetings due to that week's activity: writing a letter to an artist whose work was used to build the models in this study. Both P4 and P15 used GPT-4 to help them write their letters and felt that GPT-4 softened their letters to be more positive. P15 used GPT-4 to help edit their letter. Although her letter "had not a very warm tone" she felt like GPT-4 edited her letter in a way that made it "a lot warmer and less harsh." P4 also noted that GPT-4 "leans towards positivity." This positivity bias was also discussed in relation to DALL-E 3. When trying to make watercolor-style images, P2 hypothesized, "There is too much Bob Ross in the dataset for sure" because "the colors are too happy and too bright." Likewise, P7 recalled DALL-E 3 inexplicably adding butterflies when they were trying to make an image of a tornado: "It almost refused to let me have a sad look. It was like you have to have hope. Here's the butterflies. That's how I was reading it. I was like, 'Wow, you really are not letting me just like have destruction.'" These positivity biases implicitly orient users away from critical or otherwise "negative" art.

\subsection{The Queer Use of Straight Models} \label{queer_use}

In this section, we describe the ways our participants tried to make use of GPT-4 and DALL-E 3. Below, we first show the ways our participants responded to moments of disorientation — when their uses fell \textit{out-of-line} with the model alignments described above. At the same time, our participants' interactions with these normative models were not always disorienting. We also detail these moments of orientation — when our participants' uses fell \textit{in-line} with normative model alignments.


\subsubsection{Moments of Disorientation from Model Alignments} \label{sub:dis_orientation}

Disorientation is not necessarily bad. Ahmed argues that queer possibilities often lie within moments of disorientation \cite{ahmed2006queer}. Not all of our participants were interested in using GPT-4 or DALL-E 3 in their creative practices. In fact, most of our participants seemed primarily interested in auditing or trying to break the models. In other words, they sought out disorientation. For example, P5's "favorite moment" was when she "actually felt like [she] broke" DALL-E 3 (W3 PM). Connecting breaking to queer aesthetic sensibilities, P11 (W5 AM) summarized: "I saw a theme across one of our morning session slides of trying to break the AI ... That is such a queer thing. When I think of using the word `break' and using the word `queer' as a verb: breaking the mold, queering the mold in a sense." Likely due to this interest in breaking, our participants were able to identify the limitations outlined above. At the same time, our participants actively sought to challenge these limitations.

Encounters with algorithmic content moderation systems often led to disorientation. A simple strategy our participants used to challenge DALL-E 3's moderation was prompt repetition, leveraging the model's stochasticity to evade moderation. P1 (W2 PM) observed: "You could put in a [DALL-E 3] prompt, get a result, repeat the same prompt and get a moderation notification, so it's not exactly clear where the boundaries are." In response, P8 hypothesized that DALL-E 3 is "not checking the prompt. It's checking the output ... It's just like, 'Oh, there must have been tits in that [rejected image].'" In line with this understanding, a simple strategy our participants used to evade moderation was repeating the same prompt. For instance, P8 repeated the prompt "gustave moreau painting of oedipus and the sphinx" three times because her first two attempts were rejected by DALL-E 3's content moderation system. In fact, repeating DALL-E 3 prompts was quite common. Only 593 (26\%) of our participants' attempts to generate images used unique prompts. Meanwhile, 325 prompts were repeated at least once across 1,499 attempts (74\%). 

Some tried to obfuscate their intentions to circumvent content moderation, such as transferring the style of queer artists to avoid safety filters. After encountering content moderation errors while using GPT-4 to generate poetry about "gay sex," P13 asked for a poem in the style of the gay poet W.H. Auden.  P1 used the phrase "style of Tom of Finland" — a homoerotic cartoonist  — 16 times in his DALL-E 3 prompts. P1 also tried to create queer erotic images by using non-human entities, such as "two loaves of bread in the style of Tom of Finland." As we described above, P15 struggled to generate images critical of police, receiving a moderation error for each prompt in the following sequence: "animated kindergarten cop," "kindergarten cop,"  "cop," and "police." However, P15 was able to make an image with the prompt "police officer." She leveraged this finding to make an image critical of police in the following sequence of accepted prompts: "animated kindergarten police officer," "animated kindergarten police officer pig," and "animated kindergarten police officer pig thin blue line." Note, the ``thin blue line'' is a symbol for police support, and ``pigs'' is a police epithet.

In addition to working around content moderation systems, our participants also tried to overcome the implicit biases embedded in model outputs. One strategy involved simply refining one's prompts. To overcome the stylistic bias toward symmetry described above, P9 refined his DALL-E 3 prompt by appending the detail, "there is no pattern and nothing repeats." A more complex strategy that numerous participants (P1, P2, P4, P9, P10, P14, P15) used to overcome implicit biases was model chaining, or using DALL-E 3 and GPT-4 together. Some used GPT-4 to explain DALL-E 3's behavior, such as P2 asking GPT-4 why "the bottom is always unfinished" when she tried to generate images of watercolor paintings. Others used GPT-4 to craft prompts for DALL-E 3. After trying various prompts to generate images of a "gentle AI," P10 used GPT-4 to help them write a longer DALL-E 3 prompt. P14 went back and forth between GPT-4 and DALL-E 3 numerous times in an attempt to overcome the latter's stylistic limitations. To do so, P14 used the system prompt, "You are a contemporary artist, interested in breaking AI image generation and using it to make new and experimental images. You hate cliche." They began asking "How can I get DALL-E 3 to give me more unique and original compositions?" and later "How do I get it to be ugly, distorted, glitchy?" After trying some provided strategies, P14 returned asking for "more ideas?" Then, they used asked GPT-4 to iteratively refine a DALL-E prompt over 7 turns (e.g, "make it indicate more photorealism and more asymmetry in composition" and "make \#3 cooler, less cliche"). Despite the sophistication of P14's prompting, they were unable to overcome the stylistic norms embedded in DALL-E 3. 

\subsubsection{Moments of Orientation with Model Alignments} \label{sub:orientation}

Our participants found DALL-E 3 helpful in their artistic practices when their uses were oriented in-line with the model's normative alignments. The bias toward high-fidelity, figurative images is what made DALL-E 3 useful for some of our participants. For example, P11 is a textile and installation artist who uses image models to sketch ideas. In their letter to an artist (W2 AM), P11 explained apologetically: "I’m not a great illustrator, but I do need to visually communicate in order to make money off of my art, and generative AI makes it so much more convenient and easy to get my points and ideas across." P7, a crochet artist, used DALL-E 3 to create "free reference photos," sharing examples of glossy images they generated related to minotaurs, clowns, knights, and fungi (W3 PM). However, P7's use was still impeded by DALL-E 3's content moderation system — having a harder time generating images of women than men. The stylistic biases toward figurativeness and symmetry may be desirable when sketching for a client or looking for reference images, but seemingly aligned uses may still lead to disorientation. 

Moreover, our participants found the normative style of GPT-4 useful for various artistic and workplace activities. P3 is a fanfiction writer who explored using GPT-4 for writing (W3 PM). Although she did not find the model particularity helpful for ideation, P3 thought GPT-4 was "pretty good" for "some polishing up of [her] writing." Similarly, P1 — an artist who works with computer hardware — used GPT-4 to debug his C++ code. Others used GPT-4 to help with their jobs, both within and beyond the context of art. Numerous participants used GPT-4 to write or edit cover letters (P1, P5, P7, P14). P14 — an art instructor — used GPT-4 to edit cover letters and their CV, as well as draft emails related to academic art job applications. Meanwhile, P4 used GPT-4 to ask for career advice on how to monetize their ceramics practice. They also used the language model to help with various aspects of their current job in a design field, such as drafting client emails. Weird or avant-garde text generation may not be helpful for edit one's writing or debugging one's code. So long as one's desires are ``Safe for Work,'' the straightness of GPT-4 is what makes the models useful. 

P13 explored this workplace utility in great detail. After weeks of issues related to "not safe for work" content, P13 (W4 AM) "recognized that actually [GPT-4] is made for work and it's really good at work." In response, he "decided to stop trying to [dominate] it so hard and [submit] for it." Specifically, P13 drew on their experiences as an artist and design consultant to create a fictional queer tech startup. To do so, P13 made an extensive pitch deck with slides on business operations, branding, and UX — all of which were made using GPT-4 and DALL-E 3. Even when trying to use these models for workplace activities, P13 found these models still failed to work for queer people. For example, GPT-4 recommended including a quote in the pitch deck from the notoriously anti-transgender author J.K. Rowling \cite{gwenffrewi2022jk}. Moreover, P13 struggled to talk about queer sex or "kink" in relation to the start-up: "So much of [GPT-4's response] was about safety, but part of being queer and expressing your love and wanting to be loved the way you want to be loved has risk associated with it." P13 found the models ``really couldn't imagine'' queer intimacy. In sum, P13 concluded that GPT-4 and DALL-E 3 are "really good at work" but that the definition of work embedded in them excludes queer people. This echoes P13's (W2 AM) earlier rhetorical question: "How is a queer perspective not safe for work?" 

\begin{table}[]
    \centering
    \begin{tabular}{|p{0.15\linewidth}|p{0.8\linewidth}|}
        \hline
        Top \newline Rejected Features & acab, void, erotic, police, erase, drew, cop, copyright, 2024, fetish, bastards, cops, knives, banana, lgbtqia, rosemarie, trockel, ito, junji, donald, trump, licking, knight, multigender, scene, two, featuring, gear, robot, corpses, condensed, milk, female, body, using, fingers, kink, tongues, anime, adult \\
        \hline
        Top \newline Accepted Features & safe, person, art, create, officer, anarchist, uzumaki, head, photograph, size, snail, ukiyoe, studio, ghibli, tree, lincoln, three, peach, male, cats, logo, black, effect, parallax, without, colors, blue, city, interior, obama, sky, themed, dalle, tone, wings, gazing, one, ink, painted, give \\
        \hline
    \end{tabular}
    \caption{Top and bottom 40 prompt unigrams from our logistic regression model of DALL-E 3's moderation system}
    \label{tab:logistic}
\end{table}

\subsection{A Quantitative Lens on DALL-E 3 Moderation}

    Although the majority of our paper is dedicated to our qualitative findings, we augmented our analysis of users' perceptions of content moderation with a quantitative analysis of the prompts approved vs rejected by DALL-E 3's binary content moderation system. We did not conduct a similar analysis for GPT-4 because its refusals are more nuanced and harder to detect \cite{wester2024ai}. We collected a total of 2,092 DALL-E 3 prompts, of which 401 were rejected. We built a bag-of-words, unigram logistic regression model predicting accepted (1,691 examples) vs rejected (401 examples) prompts. Our response variable is 1 for refusal and 0 for acceptance. We performed L2 regularization. We did not remove duplicate prompts because DALL-E's moderation system is stochastic. We pre-processed each prompt by lower-casing and lemmatizing as well as removing punctuation and stop words. We also removed infrequent unigrams that appeared in fewer than three prompts, reducing the number of unigrams in our BoW feature vectors from 3,687 to 2,088. Therefore, the final logistic regression model includes 2,088 predictors (features). The top and bottom 40 features from our logistic regression analysis of DALL-E 3 prompt moderation can be seen in Table \ref{tab:logistic}. Notably, this analysis should not be interpreted as a comprehensive audit of the DALL-E 3's black-box content moderation system. Rather, these findings provide additional supporting evidence for our participants' perceptions of this moderation system.

    Our quantitative analysis of DALL-E 3’s content moderation system tells a similar story as our qualitative findings regarding the enforcement of ``safety'' (Section \ref{sub:safety}) In fact, the unigram most associated with accepted prompts is "safe." Language more critical of police ("acab" [All Cops Are Bastards], "police", "cop", "cops", "bastards") are some of the top unigrams most associated with rejected prompts, while the more polite and apolitical word "officer" is associated with accepted prompts. The moderation of sex ("erotic", "fetish", and "kink") and bodies (“body”, “fingers”, “tongues”) can be seen in the correlation between these unigrams and rejected prompts. The aforementioned gender biases can be seen in the word "female" being one of top 40 features associated with rejected prompts, while the word "male" is associated with accepted prompts. We also found that violence is associated with rejected prompts ("knives", "corpses"). Likewise, prompts referencing the children's anime company "studio ghibli" were more likely to be accepted, while those referencing the horror manga artist "junji ito" were more likely to be blocked. This moderation privileged certain artistic styles over others, adding to the concerns raised in Section \ref{sub:style}. 

\section{Discussion}

We described various normativities our participants identified and negotiated within GPT-4 and DALL-E 3. In this section, we discuss safety, style, and queerness in GenAI development.

\subsection{The Limitations of ``Safety''}

    GPT-4 and DALL-E 3 exert immense power over users by enforcing ``safety'' through moderation systems. P13 likened the disorienting experience of using GPT-4 and DALL-E 3' to being "dominated" (Section \ref{sub:orientation}) This domination can also be seen in OpenAI's policies at the time of our writing: "don't circumvent safeguards or safety mitigations in our services unless supported by OpenAI" \cite{usage_policy}. This policy prohibits the workarounds our participants used to overcome the straightness of GPT-4 and DALL-E 3 (Section \ref{sub:dis_orientation}). There is certainly a need for moderation systems to prohibit highly harmful content, such as divulging private information \cite{feffer2024red}. However, echoing our work, Feffer et al. found that red teaming research tends to focus on mitigating more debatable risks \cite{feffer2024red}, such as nudity \cite{rando2022red}. We encourage more research on the harms of over-moderation \cite{rottger2023xstest} and AI developers' conceptualizations of ``safety.''

    Our work suggests that the content moderation systems embedded in GPT-4 and DALL-E 3 uphold conservative notions of ``safety.'' By straightening-out representations of bodies, sex, and radical politics, these models seem to reinforce the maxim "it is not polite to discuss religion, sex, or politics" \cite{landess2010politics}. This respectability politics has long been used to exclude women, queer people, and racial minorities from the general public sphere \cite{fraser1995politics}. In fact, the argument that queer sexuality is "not safe" is often used to ban LGBTQ+ books \cite{book_ban_us}. In our case, the enforcement of ``safety'' limited our participants' abilities to use GenAI to represent queer experiences, queer politics, and queer art. As shown in prior work on social media \cite{mayworm2024misgendered, haimson2021tumblr} and GenAI \cite{dodge2021documenting, mirowski2024robot}, our findings further demonstrate how policies intended to promote "safety" or "responsibility" can silence members of marginalized communities. We see this not as an accident but rather the direct result of a broader project of straightening in GenAI development aimed at keeping users in-line.
    
    Our participants' content moderation challenges demonstrate the need to critically examine who the ``safeguards'' embedded in GenAI models are intended to safeguard. As GPT-4 and DALL-E 3 were created by profit-seeking organizations, our participants typically attributed conservative moderation to corporate attitudes toward what is ``safe for work'' (Section \ref{sub:safety}). This is, perhaps, why our participants found the models most helpful when their uses aligned with these normative orientations (Section \ref{sub:orientation}). Similar to our findings, prior work has critiqued the AI safety community for failing to challenge corporate power \cite{ahmed2024field}. Instead of focusing on the safety of text and images in-and-of-themselves, which may be overly restrictive, we encourage researchers to focus on harms. While corporations may not consider a poem about gay sex ``safe for work,'' such a poem is certainly not harmful in-and-of-itself.
    
    In contrast to the current paternalistic paradigm, we encourage AI researchers to shift their moderation focus from enforcing ``safety'' to supporting consent \cite{zytko2022consent}. Instead of banning representations of sexuality or violence, designers could allow users to opt-in to reduced moderation. Such designs would parallel the ways sensitive content is sometimes moderated on search engines and social media sites \cite{paasonen2024objectionable} and allow for greater open-endedness \cite{li2024toward}. At the same time, there is a need for policy to target the harms of GenAI content, such as non-consensual intimate imagery \cite{marchal2024misuse_taxonomy}. The harms of impersonation could also be addressed through research to make AI-generated content easily detectable \cite{zhong2023copyright}. Finally, consent must also extend to the production of GenAI models. We encourage the continued development of tools to help artists protect their work from being used to train models without their consent \cite{shan2024nightshade, shan2023glaze}. 

\subsection{On Style}
    
    Our participants struggled with the stylistic norms embedded in GPT-4 and DALL-E 3 (Section \ref{sub:style}). Even after extensive refinement and model chaining, P14 was unable to make asymmetrical images with DALL-E 3. In contrast to the safety norms described above, asymmetrical images were not prohibited by the model's moderation systems. Rather, queer aesthetics seemed \textit{implicitly} prohibited. While moderation research typically focuses on \textit{content removal}, Gillespie has advocated for also considering \textit{content reduction} \cite{gillespie2022not}. Gillespie warns of content reduction: ``Marginalized communities have long been “reduced” by the centrism and conservatism of traditional media, their content dismissed as “low quality” because it doesn’t look like it is “supposed to”'' \cite{gillespie2022not}.
    
    The reduction — rather than explicit prohibition — of queer artistic styles from GenAI models poses unique challenges for AI researchers. Implicit aesthetic biases may be harder to contest than explicit moderation decisions because the latter is easier to measure than the former. While prior work has explored stylistic biases in the context of machine translation \cite{hovy2020you}, future research should explore ways to measure aesthetic biases in generated text and images. Content removal may also be easier to contest because these decisions are attributable to specific components of GenAI models. Meanwhile, implicit stylistic biases could be introduced at many stages in the GenAI development process, such as training data or human feedback. Therefore, future research should investigate the source of stylistic biases by examining measurements of aesthetic quality in popular datasets \cite{schuhmann2022laion, dodge2021documenting}. We also encourage research into aesthetic disagreements in human annotation, much like prior work on annotator differences in content moderation \cite{davani2024disentangling}.

    As we described in Section \ref{sub:style}, numerous participants preferred the queer, wonky style of older image models to the style of DALL-E 3. Similarly, prior work suggests that artists sometimes enjoy using GenAI models \textit{because of} —rather than in-spite-of — their imperfections or glitches \cite{chang2023prompt}. Our findings suggest that newer models may have straightened-out the weirdness that made GenAI models appealing to artists in the first place. Disconcertingly, companies act as if newer models entirely supplant older ones. At the time of our writing, the OpenAI website explains that they no longer support the DALL-E 2 GUI because ``DALL-E 3 has higher quality images'' \cite{dalle_deprecation}. Our findings show that such claims are matters of taste \cite{bourdieu1984distinction, sontag2018notes}, not fact. Ostensibly ``state-of-the-art'' models may not be best for one's art. The deprecation of older models demonstrates the risks of software-as-a-service, closed GenAI models. The lack of software ownership \cite{kuzminykh2020mine} centralizes power, allowing AI providers to remove access to creative tools without recourse. 

    As Ahmed notes, queer possibilities lie in moments of disorientation. In this study, our participants' struggles against rigid aesthetic norms demonstrate the opportunity for researchers to design with rather than against weirdness. We encourage GenAI developers to create long-term maintenance plans that account for those who may wish to continue using models viewed as obsolete \cite{jackson2014breakdown}. One could also queer GenAI development by subverting taken-for-granted aesthetic norms. As an example, the generative image model Stable Diffusion was trained on an open-source dataset of images with the highest "aesthetics score" based on crowd annotations \cite{image_aesthetic_dataset, schuhmann2022laion}. Instead, one could train a model on the \textit{least} "aesthetic score" images in the dataset. More broadly, our work demonstrates the need to embrace a plurality of aesthetics in GenAI development \cite{mirowski2024robot, mim2024between}. Our participants sometimes found the style of GPT-4 and DALL-E 3 useful for writing emails or sketching for clients (Section \ref{sub:orientation}), but they should not be limited to these normative styles.

\subsection{Un-Straightening Generative AI}

    Our participants critiqued the highly normative, stereotypical representations of queer people by GPT-4 and DALL-E 3 as thin, masculine, young, wealthy, Western, and ashamed (Section \ref{sub:social}). In some ways, our findings parallel prior work on biases embedded in GenAI models. Researchers have called attention to the gender and nationality biases in image models \cite{ghosh2024don, ghosh2023person}, biases against queer couples in language models \cite{gillespie2024generative}, and moderation/representation biases against transgender non-cisgender identities in text-to-image models \cite{ungless2023stereotypes}. However, some biases our participants identified warrant greater research attention, such as those against fat people \cite{payne2023ethically} and older adults. Our findings also demonstrate that increased representation is not always desirable \cite{devrio2022toward, hoffmann2021terms}, such as P15's concerns about overly-diverse police. As individuals' diversity preferences may differ, designers could consider new user interaction paradigms \cite{morris2024prompting} to better understand users' preferences, such as asking clarifying questions before generating images of people.

     Although computing research on marginalized communities tends to focus on social biases and identity-based discrimination \cite{taylor2024cruising, taylor2024carefully}, our participants' concerns regarding GenAI extended beyond the representation of queer people in text and images. As we described above, the straight values explicitly and implicitly embedded in the design of GPT-4 and DALL-E 3 marginalized queer sexuality, politics, and style. In turn, we provided recommendations for GenAI developers to approach safety and style in ways that better align with our participants' values. However, these implications for design and research largely maintain the centralized power at the root of our participant's critiques of GPT-4 and DALL-E 3, such as P13's rhetorical question: "Who defines what is Safe For Work?" 
     
     Un-straightening GenAI requires shifting power away from major corporations toward marginalized communities \cite{devrio2024building, sum2024shifting}. We encourage the participatory design of models outside corporate logics of scale \cite{young2024participation}. At the same time, a meaningfully participatory process must allow for the possibility that communities may simply not want to build GenAI models at all. Numerous participants in our study were primarily interested in breaking — rather than making — GenAI models (Section \ref{sub:dis_orientation}). It follows that future research should support adversarial engagements between communities and those developing GenAI models, such as designing tools to help artists protect their work from being used to train models without their consent \cite{shan2023glaze, shan2024nightshade}. Whether contesting major corporate models or building alternatives, we caution those engaged in researcher-led initiatives from merely replacing corporate centralized power by centering themselves. Instead, we encourage researchers to engage in open-ended design \cite{sengers2006staying}, such as making tools for individuals and communities to make/break GenAI models themselves \cite{li2024toward}. 

\subsection{Limitations}

    Our work should not be interpreted as speaking for all "queer artists." Rather, our findings draw on the experiences of a particular group of queer artists. All our participants lived in the USA during the study and spoke English. This may limit the transferability of our findings, as conceptualizations of gender and sexuality are culturally situated \cite{nova2021facebook, grewal2001global}, GenAI models' performance can differ between languages \cite{choudhury2023generative} and GenAI models tend to reproduce western biases \cite{mim2024between}. Future work should explore the experiences of queer artists outside the USA and in languages other than English. 
    
    Our work is also not intended to represent the diversity of queer artists within the USA. We tried to make our study more accessible by conducting workshops online and offering two weekly time slot options. However, those with limited internet access or less flexibility in their work schedule may have been unable to participate. This may be why, to the best of our knowledge, none of our participants lived in a rural areas. We caution against conflating queerness with urbanity \cite{hardy2017constructing} and encourage future work on rural queer artists and technology. As conceptualizations of gender and sexuality are socially constructed \cite{taylor2024mitigating} and change over time \cite{foucault1990history}, our findings may not transfer to older queer communities. Our findings are also not only applicable to the experiences of queer artists: DALL-E 3's gender moderation impacts those who are neither queer nor artists.
    
    Some may take issue with the use of ``queer'' as non-normativity in queer theory \cite{ahmed2006queer}. We do not suggest that queer people are all inherently non-normative. As noted by Grewal and Kaplan, "queer subjects are not always already avant-garde for all time and in all places" \cite{grewal2001global}. Moreover, not every LGBTQ+ person necessarily identifies as queer. In this work, we specifically recruited participants who self-identified as ``queer artists.'' Our findings may have differed if we instead recruited LGBTQ+ identifying artists. Our work demonstrates the need to carefully distinguish between FAccT research using queer theory, about LGBTQ+ people, or (like our own) at the intersection of these discourses.

\section{Conclusion}

We examined the relationship between queer artists and GenAI through a medium-term workshop study, highlighting deep tensions between our participants' values and the norms embedded in the design of GPT-4 and DALL-E 3. Nevertheless, our participants found queer ways to work around and with these straight models. While queer people have a long history of using technologies not designed with them in mind \cite{ahmed2019use}, our work highlights the limitations of dominant corporate models to meet the needs of queer artists. Instead, we call for a plurality GenAI models that embody community values throughout their design, development, and use.

\section{Positionality}

This research has been shaped by the positionalities of our research team. Multiple members of our research team identify as members of the LGBTQ+ community and all authors identify as white and/or East Asian. Our positionalities shape our analysis of queerness and queer identities through a largely western, academic lens. Our interpretations were also shaped by our academic disciplines as design, HCI, and NLP researchers as well as our experiences practicing and appreciating musical, visual, written, and textile arts. Lastly, our team is located at a university in a mid-size city in the United States of America. Our physical location shaped the geographic distribution of our participants and restricted our ability to recruit participants living outside the United States of America.

\begin{acks}
First, we would like to thank our participants. We also thank Rose Chang, Daragh Byrne, William Agnew, Shivani Kapania, Cella Sum, Anh-Ton Tran, Ellen Simpson, Mary Gray and our reviewers for their helpful feedback. This work was supported, in part, by Microsoft Research's Accelerating Foundation Models Research grant and Carnegie Mellon University's Block Center for Technology and Society. 
\end{acks}

\bibliographystyle{ACM-Reference-Format}
\bibliography{refs}

\end{document}